\documentclass[aps,prb,twocolumn,superscriptaddress,floatfix]{revtex4}
\usepackage{graphicx}
\begin{document}


\title{High-Temperature Hall Effect in Ga$_{1-x}$Mn$_x$As}


\author{D. Ruzmetov}
\author{J. Scherschligt}
\author{David V. Baxter}
\email[E-mail address: ]{baxterd@indiana.edu}
\affiliation{Dept. of Physics, Indiana University, Bloomington, IN, 47405, USA}
\author{T. Wojtowicz}
\affiliation{Dept. of Physics, University of Notre Dame, Notre Dame, IN, 46556, USA}
\affiliation{Institute of Physics, Polish Academy of Science, aleja Lotnikow 32/46, 02-668, Warsaw, Poland }
\author{X. Liu}
\author{Y. Sasaki}
\author{J.K. Furdyna}
\affiliation{Dept. of Physics, University of Notre Dame, Notre Dame, IN, 46556, USA}
\author{K.M. Yu}
\author{W. Walukiewicz}
\affiliation{Materials Sciences Division, Lawrence Berkeley National Laboratory, Berkeley, CA, 94720, USA}


\date{\today}

\begin{abstract}

The temperature dependence of the Hall coefficient of a series of
ferromagnetic Ga$_{1-x}$Mn$_x$As samples is measured in the
temperature range $80\,{\mathrm K} <T<500\,{\mathrm K}$. We model the
Hall coefficient assuming a magnetic susceptibility given by the
Curie-Weiss law, a spontaneous Hall coefficient proportional to
$\rho_{xx}^2(T)$, and including a constant diamagnetic contribution in
the susceptibility. For all low resistivity samples this model
provides excellent fits to the measured data up to $T$=380\,K and
allows extraction of the hole concentration ($p$). The calculated $p$
are compared to alternative methods of determining hole densities in
these materials: pulsed high magnetic field (up to 55\,Tesla)
technique at low temperatures (less than the Curie temperature), and
electrochemical capacitance- voltage profiling. We find that the
Anomalous Hall Effect (AHE) contribution to $\rho _{xy}$ is substantial even
well above the Curie temperature.  Measurements of the Hall effect in 
this temperature regime can be used as a testing ground for theoretical
descriptions of transport in these materials. We find that our data are
consistent with recently published theories of the AHE,
but they are inconsistent with theoretical models previously used to 
describe the AHE in conventional magnetic materials.
\end{abstract}


\maketitle

\section{Introduction}
Random alloys of GaAs and Mn have attracted interest from a number
of research groups \cite{dietl01,edmonds,y_ohno99,konig01}. Over a
limited composition range Ga$_{1-x}$Mn$_x$As is ferromagnetic with
a Curie temperature as high as 140\,K \cite{edmonds} and Mn content
$x$ up to 0.1. This class of III-V diluted magnetic semiconductors
(DMS) are fabricated by means of low temperature molecular beam
epitaxy (MBE) which insures that most of Mn atoms randomly
substitute cations in their crystal positions. They are promising
materials in spintronics due to their potential for tuning their
ferromagnetic and electric properties by means of electrical field
\cite{h_ohno00}, optical excitation \cite{Koshihara}, or
impurities \cite{Satoh01}. Ferromagnetic GaMnAs can be grown
isomorphically within semiconductor structures and may be suitable
for spin injection into GaAs based devices \cite{y_ohno99}.

A complete theory of ferromagnetism in GaMnAs has yet to be
established, but there is general agreement that a ferromagnetic
coupling between local moments on the Mn ions is mediated by
holes moving through the lattice.  A variety of different theoretical 
frameworks have been put forward, ranging from a mean-field approach 
with essentially free carriers\cite{dietl01, konig01} to approaches 
that assume the carrier dynamics to be dominated by the
proximity of the system to the metal insulator transition\cite{Berciu01}
and the importance of disorder on the magnetic state\cite{DasSarma03}. 
To date many of the claims regarding the relevance of these theories have
been based upon their ability to predict the ferromagnetic
transition temperature. In trying to distinguish between these
many competing theories it is essential to explore as wide a range
of physical properties as possible. Noting that these alloys span
a wide range of conductivities, measurements of electronic transport
could be particularly useful in this regard.

Since it is generally agreed that holes mediate the ferromagnetic 
interaction between Mn ions the hole concentration is a crucial 
parameter for determining the properties of these materials. It
influences essentially all their major properties \cite{dietl01,
konig01}. Therefore it is also important to measure the carrier density
and understand the physical factors that control it in order to
improve our knowledge of these materials. However, the disorder in
GaMnAs results in a carrier concentration ($p$) that is generally
lower than the Mn density making  an independent measurement of
$p$ necessary. Determination of the carrier density by means of
the Hall effect is complicated due to the presence of the
Anomalous Hall contribution arising from the broken symmetry
provided by the magnetisation. A conventional way to solve this
problem is going to low temperatures and high (above 20\,Tesla)
magnetic fields where the magnetization and magnetoresistance
saturate making possible to extract the Ordinary Hall coefficient
\cite{omiya, baxter02}. Limited accessibility of this method has
stimulated the search for other techniques of determining $p$
\cite{Yu02, Seong02}.

In this paper we measure the Hall effect in a series of
Ga$_{1-x}$Mn$_{x}$As samples at temperatures above
$T_c$ in order to provide insight on both of these fronts (testing
our understanding and providing a means for measuring the carrier 
concentration). Our results provide support for the recent
theory of Jungwirth \textit{et al}\cite{Jungwirth02, Jungwirth03} 
suggesting that a clean-limit theory can successfully describe 
transport in these materials over a range of
compositions. We also demonstrate that Hall measurements in this
temperature range can be used to find the hole concentration
without the need to resort to high magnetic fields. The
temperature dependence of the Hall coefficient in this
paramagnetic regime may be described by a model that accounts for
the paramagnetic contribution to the Anomalous Hall Effect (AHE)
together with the Ordinary Hall Effect (OHE). With this model we
demonstrate that the AHE can dominate the Hall resistivity in GaMnAs
even above room temperature. However, by properly accounting for
this contribution the carrier concentration can be determined from
the measurements.

\section{Experiment}

GaMnAs films were grown by means of MBE at a substrate temperature
$T_s=275^{\circ }$C. The sample structures were as follows:
semi-insulating GaAs (100) substrate/100\,nm of GaAs deposited by
MBE at $T_s=590^{\circ }$C/30--100\,nm of GaAs deposited by MBE at
$T_s=275^{\circ } $C/GaMnAs film with thicknesses of 123\,nm or
300\,nm. All films demonstrate an in-plane easy axis of the
magnetization. The samples were patterned for transport
measurements in a standard Hall-bar geometry with 
dimensions between the longitudinal voltage contacts of
$0.9 \times 0.3 $\,mm$^2$. For High-Temperature Hall and resistivity
measurements we used an MMR Technologies R2105-26 Thermal Stage
System employing Joule-Thomson Effect to vary sample temperatures
in the range $80-500$\,K. The samples were mounted on the thermal
stage with silicon grease and were kept in $\sim 15$\,mTorr vacuum
during measurements. The thermal stage was placed between the
poles of an electromagnet so that a DC magnetic field (up to 5\,kOe) 
could be directed perpendicular to the sample plane. The 
samples were wired with golden wires and soldered with Indium. The
contacts worked even above the In melting point (430\,K) since the
wires were held on the sample by surface tension of liquid In.

Transport measurements above 80\,K were done using standard Lock-in
techniques with excitation current of $100\,\mu$A and frequency
$17$\,Hz and also some of the results were confirmed with DC
measurements using Keithley  current source 220 and
nano-voltmeter 182. Each Hall coefficient data point was obtained
from the slope of a straight line fitted to at least 6 data points
of the Hall resistance vs. field curve for 
a magnetic field ranging from -5\,kOe to +5\,kOe.
We also did some measurements in
van der Pauw geometry which agreed with Hall-bar measurements. The
low temperature (below 80\,K) transport measurements were performed
in a liquid He cryostat using Quantum Design digital bridge model
1802 with $50\,\mu$A excitation current. 
The Hall coefficients below the Curie temperature were measured at such
magnetic fields that the sample magnetization in the growth (hard) direction 
was well below the saturation, which typically corresponded to a range $-500\,\mathrm{Oe} < H < +500\,\mathrm{Oe}$.
The magnetization of
ferromagnetic GaMnAs was measured in a Quantum Design MPMS XL
Superconducting Quantum Interference Device magnetometer. Annealing
of GaMnAs samples was done in situ on the MMR thermal stage
between transport measurements. The sample was heated to
$260^{\circ }$C and kept for 2\,hrs in 15\,mTorr vacuum. Similar
samples annealed on a specially designed annealing apparatus in a
high purity Ar gas mass-controlled flow and with short ($\sim
5$\,min) heating up and cooling down times  showed the same magnetic
and transport properties as the ones annealed on the MMR stage.

\section{Results and Discussion}

\begin{figure}[h]
\includegraphics[width=3.5in]{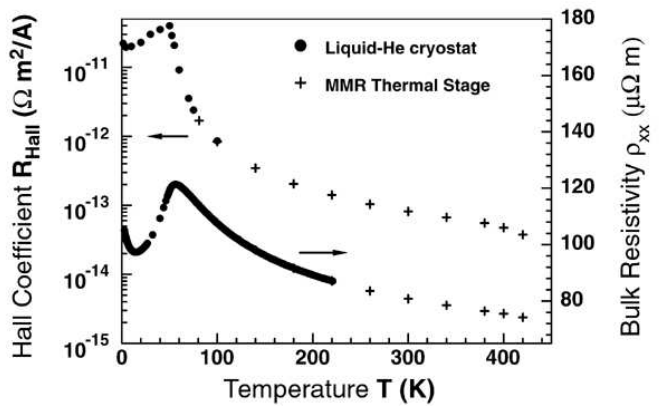}
\caption[R_Hall_ro]{Hall coefficient and resistivity of GaMnAs
($x=0.048$, $T_{c}$=53\,K) were measured in a wide temperature range
using Liq-He cryostat below 100\,K and MMR Joule-Thomson
refrigerator above 80\,K. 
The Hall measurements were made over a field range of $\pm 5$\,kOe for 
temperatures above $T_c$ (on the MMR system) and over a range of $\pm 500$\,Oe 
below $T_c$ (in the Liq-He cryostat). \label{R_Hall_ro}}
\end{figure}

The bulk resistivity  and the Hall coefficient:
\begin{equation}
R_{Hall}\equiv \frac{\rho _{xy}}{H_z}=\frac{E_{y}}{j_{x}H_z}
\end{equation}
where $\rho _{xy}$ is the Hall resistivity, $E$ and $H$ are
electric and magnetic fields, and $j$ is current density, were
measured in a wide temperature range $2 - 420$\,K using a
conventional LHe cryostat below 100\,K and the MMR thermal stage
above 80\,K. The temperature dependence of $R_{Hall}$ and the
resistivity for a representative sample with $x=0.048$  are shown
in Fig.~\ref{R_Hall_ro}.  The Hall coefficient and resistivity
exhibit maxima approximately at the Curie temperature ($T_c$) 53\,K.
Note that the Hall coefficient continues to change with $T$ above
300\,K, which is an indication of the presence of the temperature
dependent anomalous Hall effect.

To account for the measured $R_{Hall}$ temperature dependence we
modeled the Hall effect as a sum of the Ordinary (OHE) and
Anomalous \cite{Chien} (AHE) Hall contributions:
\begin{equation}
R_{Hall}=\frac{\mu _o}{e\cdot p}+R_s \cdot(\frac{\chi _c}{T-\theta
}+\chi _{o}) \label{RHall}
\end{equation}
The first term in Eq.~(\ref{RHall}) is the ordinary Hall coefficient
due to the Lorentz force with $\mu _o$ being magnetic permeability of
the vacuum, $e$ -- the magnitude of the electron charge, and $p$
standing for the free hole concentration. The AHE is proportional to
the magnetization of the sample and this takes the form of the  second
term in the equation in the paramagnetic regime.  $R_s=\gamma_{para}
\cdot \rho^n_{xx}(T)$ is the spontaneous Hall coefficient with
$\gamma_{para}$ and $n$ -- temperature independent parameters, and
$\rho _{xx}$ -- the bulk resistivity of the GaMnAs film. We model the
paramagnetic susceptibility following Curie-Weiss law with $\chi
_c=\mu _o N_{Mn} g^2 J(J+1) \mu ^2_B / (3k_B)$ -- the Curie constant
taking $g=2$ and $J=5/2$, and $\theta$ is the Curie-Weiss temperature.
$\chi_o$ is a temperature independent correction to the paramagnetic
susceptibility which is necessary to be included in order to
adequately describe the measured data with a single form for $R_s$.
$\chi _{o}$ is the same for all samples and always smaller than the
paramagnetic term in our fits.

\begin{figure}[h]
\includegraphics[width=3.5in]{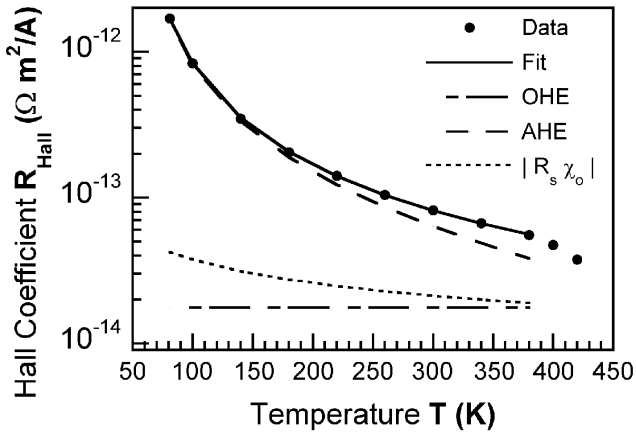}
\caption[R_Hall_fit4p8]{Fitting to the measured Hall coefficient of
the $x=0.048$ sample with $T_c$=53\,K. Fit is Eq.~(\ref{RHall}) and is the sum of AHE and OHE.  
Fitting parameters:
$p=4.5\times 10^{-20}\,\mathrm{cm}^{-3}$, $\gamma_{para} =0.033\ ({\mathrm
A}\cdot \Omega )^{-1}$, $\theta =57\,{\mathrm K}$, $\chi
_{o}=-10^{-4}$. \label{R_Hall_fit4p8}
The dotted line shows the absolute value of the contribution  to the $R_{Hall}$ due to $\chi_o$.
Note that although $\chi_o$ is itself a constant, the corresponding contribution to
the Hall coefficient varies with temperature due to the dependence of $R_s$ (in Eq.~\ref{RHall}) 
on the longitudinal resistivity.
}
\end{figure}

\begin{figure}[h]
\includegraphics[width=3in]{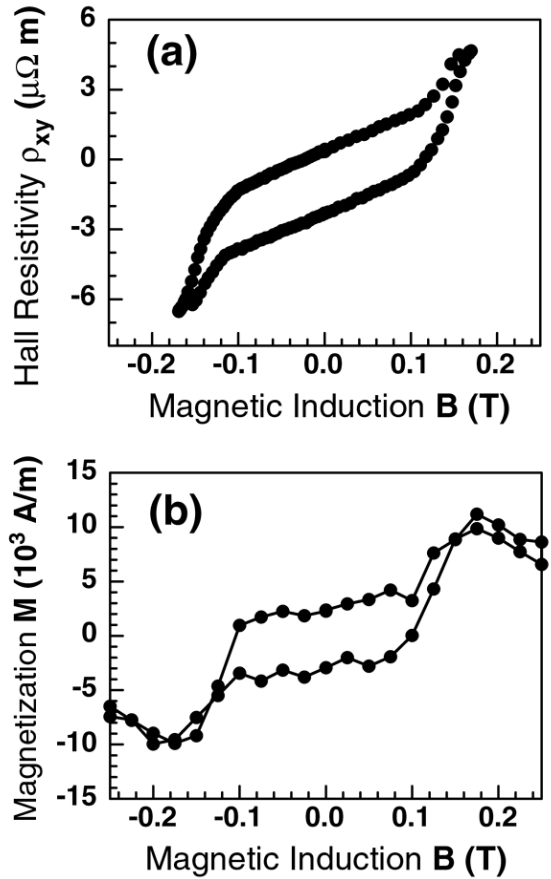}
\caption[gamma]{Measured Hall resistivity (a) and magnetization
(b) of Ga$_{1-x}$Mn$_x$As $x=0.048$ sample at 4.2\,K. Magnetic field
is in the hard axis direction. The magnetization graph shows raw
SQUID data (circles) which includes diamagnetic contribution from
the GaAs substrate. \label{gamma}}
\end{figure}

We were able to fit this model with $n=2$ to the measured data for all
samples with room temperature resistivities less than $100 \,\mu \Omega$m. Carrier concentrations were extracted from these fits. A typical
fit  is shown in Fig. \ref{R_Hall_fit4p8} along with the separate
contributions from the OHE and AHE.
In order to show the influence of the $\chi_o$ correction the figure includes the $|R_s \cdot \chi_o|$ term.
\begin{table*}[t]
\parbox{6.in}{\caption[Summary]{Comparison of different methods of
measuring the hole density for a set of Ga$_{1-x}$Mn$_x$As samples.
Calculation of $p$ using: 1) $R_{Hall}$ at 340\,K assuming that the
whole Hall voltage is due to the OHE, $p_0$; 2) high fields (above
25\,T) and low temperatures at the Los Alamos National Lab facilities,
$p_1$ (taken from ref.  [{\large $_{\cite{baxter02}}$}]); 3)
electro-chemical capacitance voltage profiling at the Lawrence
Berkeley National Lab, $p_2$; 4) the high-temperature Hall method
described in this paper, $p_3$.
  \label{Summary}}    }

\begin{center}

\begin{tabular}{*{9}{|c}|}\hline
\ $x$ \ & Thick- & $T_c$ & $\rho _{xx}(300\,{\mathrm K})$   & $\gamma_{para}$ &$p_0$
       & $p_1$   [{\large $_{\cite{baxter02}}$}]                  &
       $p_2$                      & $p_3$\\ 
 & ness &  &$\pm5\%$      & \ $\pm
       0.002$ \      &ignoring AHE at 340K       &$T\ll T_c$, $H>25\,{\mathrm T}$
       &ECV                         & $T_c<T<380\,{\mathrm K}$\\ 
& (nm)  &(K)    &$(\mu
       \Omega {\mathrm m})$& $({\mathrm A}\Omega)^{-1}$&($\times 10^{20}\,\mathrm{cm}^{-3}$)
       &($\times 10^{20}\,\mathrm{cm}^{-3}$) &($\times 10^{20}\,\mathrm{cm}^{-3}$)  &
       ($\times 10^{20}\,\mathrm{cm}^{-3}$)\\ \hline
\hspace{0.1in}0.033\hspace{0.1in}
       & 300  &41     &77     &   0.039      &$1.4\pm 0.1$        & $3.7\pm 0.7$              & $2.8 \pm 0.6$              & $2.8\pm 0.3$\\
0.048  & 300  &53     &81     &   0.033      &$1.2\pm 0.1$        & $3.0\pm 0.6$              & $4.4 \pm 0.3$              & $4.5\pm 0.7$\\
0.050  & 300  &55     &60     &   0.044      &$1.4\pm 0.1$        & $4.5\pm 0.9$              & $3.2 \pm 0.6$              & $4.4\pm 0.8$\\
0.053  & 300  &55     &92     &   0.026      &$1.0\pm 0.1$        & $2.3\pm 0.5$              & $3.5 \pm 0.3$              & $4.0\pm 0.7$\\
0.055  & 300  &59     &69     &   0.041      &$1.1\pm 0.1$       & --                        & $4.5 \pm 0.6$              & $4.2 \pm 0.7$\\
0.072  & 124  &63     &127    &    --        &$0.45\pm 0.06$      & --                        & $4.0 \pm 0.3$              & $3^{+2}_{-1}$
 \footnote{ $n=1.4$ ($\rho _{xx}$ exponent) only in this fit;}\\
0.072  & 124  &84 \footnote{Annealed at $260^{\circ}$C for 2\,hrs.}
               &79     &   0.031      &$0.80\pm 0.06$             & --                        & $6.8 \pm 0.7$              & $6\pm 2$\\
\hline
\end{tabular}
\end{center}
\end{table*}
One can see that our model perfectly fits the data up to 380\,K. For
this sample the AHE contribution dominates over OHE in the whole
temperature range of the fit. The fitted Curie-Weiss temperature
is slightly bigger than $T_c$ which is a common thing in
ferromagnetic metals \cite{Martin1}. Since $\chi _{o}$ was held
constant for all samples, $p$, $\theta$, and $\gamma_{para}$ were
essentially the only free parameters in the fit for each sample.
The Curie-Weiss temperature $\theta$ was allowed to vary but the
optimal value was always within 4\,K of $T_c$ determined from SQUID
magnetometer measurements. To confirm the reliability of our
method we estimate $\gamma_{para}$ from independent measurements
of the AHE at 4.2\,K using the relation $\rho _{xy}=\gamma_{ferro}
\cdot \rho ^2_{xx}  M$, since at this temperature an independent
measurement of $M$ is possible. Hall $(\rho _{xy})$ and bulk
$(\rho _{xx}=103\pm 2 \, \mu \Omega  \mathrm{m})$ resistivities were
measured in the LHe cryostat at 4.2\,K (Fig.~\ref{gamma}a) and
magnetization $M$ was found in the MPMS (Fig.~\ref{gamma}b).
The magnetic field was in the hard axis direction which explains
relatively large coercive force of $\sim 0.12\,$T and small
magnitude of the magnetization. In the MPMS data, the diamagnetic
contribution from the GaAs substrate causes the magnitude of the
measured magnetization to decrease above 0.18\,T where the
ferromagnetic signal saturates. But this does not affect our
measurement of $\gamma_{ferro}$ since  the remnant magnetization
and  the Hall resistivity at $B=0$ are used in the calculation. The
value so obtained, $\gamma _{ferro}=0.04 \pm 0.01\,(\mathrm{A}\Omega)^{-1}$, agrees within the uncertainties with
$\gamma_{para}$ determined from the fit  $\gamma _{para}=0.033\pm
0.002\,(\mathrm{A}\Omega)^{-1}$  in the paramagnetic region. This result supports
our model and is evidence that the relation $R_s=\gamma \cdot \rho
^2_{xx}(T)$   is valid in a wide temperature range (below and
above $T_c$).  This confirms that the
physical mechanism responsible for the AHE is the same for
ferromagnetic and paramagnetic phases of GaMnAs.

The carrier concentrations determined by our method ($p_3$) are
summarized in the Table \ref{Summary}. In determining $p_3$, the
spontaneous Hall coefficient for all low resistivity ($ \rho
_{xx}(300\,\mathrm{K}) < 100\,\mu \Omega \mathrm{m}$) GaMnAs samples was assumed to vary as
$R_s=\gamma _{para}\cdot \rho ^2_{xx}(T)$. In the  conventional
treatment of the AHE \cite{Chien}, a quadratic dependence of the Hall
coefficient on the diagonal resistivity is taken to suggest that the
side-jump is the dominant scattering mechanism in the studied
material.  However, since in this conventional picture the side jump
mechanism causes the variation $\rho_{xy} \propto \rho^2_{xx}$ and
the skew scattering varies as $\rho_{xy} \propto \rho_{xx}$, then  one
could expect $n=2$ (exponent of $\rho_{xx}$)  to be favored in
high-resistivity materials. This disagrees with our observations when
we see that $n=2$ assumption works for  low-resistivity samples and
breaks down for high-resistivity ones. A good illustration of this is
the fitting results for the  $x=0.072$ sample shown in
Table~\ref{Summary}.  The as grown sample with  $x=0.072$ has  $\rho
_{xx}(300K)=127\,\mu \Omega \mathrm{m}$ and in this case  it was necessary to
set the exponent $n=1.4$ (defined by $R_s = \gamma_{para} \cdot 
\rho^n_{xx}$) in order to achieve good fitting with Eq.~(\ref{RHall}).
The room temperature resistivity of this sample decreases to $79\,\mu
\Omega \mathrm{m}$ upon annealing and a good fit with $n=2$ becomes possible
for this annealed sample. In general, samples with room temperature
resistivity higher than $100\,\mu \Omega \mathrm{m}$ show more rapid decrease of
resistivity with increasing $T$ and their Hall data cannot be fitted
with $n=2$. Samples with $\rho _{xx}$ higher than $180\,\mu \Omega \mathrm{m}$
cannot be fitted with any $n$. These highly resistive samples have
large Mn concentrations (above 7\%) and may lie on the insulator side
of the metal-insulator transition\cite{H.Ohno99}. It is worthwhile to
note that the determination of $p$ by means of the Hall effect
measurements at high fields and low temperatures also fails for
insulating samples due to their large magnetoresistance
\cite{H.Ohno99}. We see that this limitation extends to the
high-temperature technique as well.

On the other hand, the recent theory of the AHE in ferromagnetic
semiconductors proposed by  Jungwirth {\em et al}\cite{Jungwirth02,
Jungwirth03}  assumes that AHE arises as an anomalous contribution to
the Hall conductivity $\sigma_{xy}$ which in the absence of disorder
in the material does not depend on $\rho_{xx}$.  That should result in
a Hall resistivity that is   proportional to  the second power of
$\rho_{xx}$ (from $\sigma_{xy}=-\rho_{xy}/(\rho_{xx}^2 + \rho_{xy}^2)$
using $\rho_{xy}^2 << \rho_{xx}^2$), i.e.  the theory predicts $n=2$
 for clean samples. Then our
fitting results show that the low resistivity samples, or equivalently
the samples with low Mn  concentration ($x < 0.06$), have the amount
of defects low enough so that the Hall conductivity  is  not
significantly affected by scattering and the GaMnAs crystal can be
considered to be clean from the electron transport point of view (or
at least regarding $\sigma_{xy}$).  This explains why $n=2$ law in
Eq.~(\ref{RHall}) works for Ga$_{1-x}$Mn$_x$As with $x<0.06$.  The
samples with high Mn content and high resistivities have more defects in the lattice, so that
the Hall conductivity is no longer defect (and $\rho_{xx}$)
independent and the relation $\rho_{xy} \propto \rho_{xx}^2$ is no
longer valid. Thus our ability to fit with $n=2$ only the data from
low resistivity samples is more consistent with this theoretical
framework\cite{Jungwirth02, Jungwirth03} than it is with the
conventional framework developed for metallic samples\cite{Chien}.  

In our fits, $\chi_{o}$ was taken to be
$-10^{-4}$(dimensionless in SI units) for all samples. The fact
that this parameter does not depend on temperature and Mn content
suggests that it is due to the diamagnetic contribution of GaAs
matrix, but its size is considerably larger than expected
($\chi_{GaAs}=-1.22\times 10^{-6}$). This could reflect a
difference between the efficiency of paramagnetic spins and
diamagnetic currents in contributing to the anomalous Hall
conductivity. If we modify Eq.~(\ref{RHall}) slightly by writing
$\chi_{o} = \nu\chi_{\mathrm{GaAs}}$ where $\nu$ is the ratio
$\gamma_{diam}/\gamma_{para}$, we find that $\gamma_{diam} =
82\gamma_{para}$. Alternatively, it is also possible that the
parameterization we have used covers some other deviation from the
Curie law in the paramagnetic susceptibility. Further work will be
needed to resolve this.

Carrier concentrations of our samples were estimated using other
techniques as well, as displayed in Table \ref{Summary}. For four
samples, the Hall resistance was measured at high magnetic field (up
to 55\,T in a pulse) and low temperature (down to 600\,mK), where
magnetization and magnetoresistance saturate and the slope of the Hall
resistance is due only to the OHE \cite{baxter02}. Such measurements
were performed at Los Alamos National Lab  and the results are shown
in the Table~\ref{Summary} as $p_1$. Table~\ref{Summary} also presents
results of the hole concentration measurements by means of
electrochemical capacitance-voltage (ECV) profiling, $p_2$. The ECV
method is described elsewhere \cite{Faktor} but for now we note that
it involves etching the sample which, therefore, is sacrificed during
the measurement.  The ECV profiling measurements showed little change
of the hole concentration along the growth direction of the GaMnAs
films, thus confirming that $p$ was a single-defined quantity in our samples.  One
can see that hole densities measured by the present method ($p_3$) are
within a factor of 2 of the high-field data ($p_1$) and within the
uncertainties of ECV measurements ($p_2$). This shows that our method
gives reasonable values of $p$ while being relatively simple and
non-destructive. Samples with high manganese concentration have bigger
AHE contributions. This complicates the extraction of the OHE
component and accounts for the large uncertainties for the sample with
$x=0.072$. Finally, to emphasize the importance of the AHE even at
high temperatures, Table~\ref{Summary} also includes carrier
concentrations obtained if you ignore AHE at 340\,K, $p_0$. These values
are systematically lower than any of the other independent
measurements of $p$. Therefore, as suggested in
Fig.~\ref{R_Hall_fit4p8} for one sample, the AHE is significant even
above room temperature in all the samples we measured and cannot be
neglected in  calculating the carrier densities.
\begin{figure}[b]
\includegraphics[width=230pt]{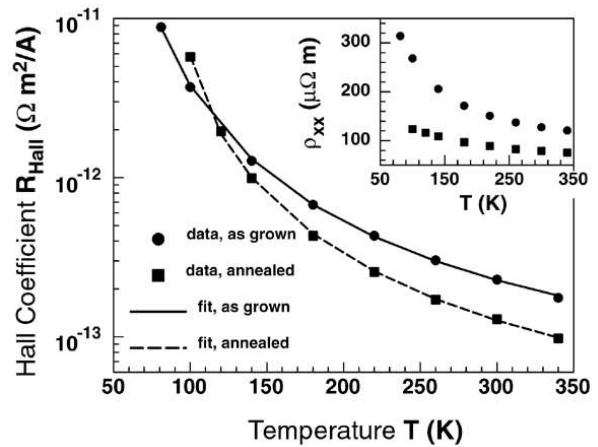}
\caption[RHall_anneal]{Annealing effect on the Hall coefficient
and bulk resistivity for $x=0.072$. The $T_c$ raises from 63\,K to
84\,K upon annealing at $262^{\circ}$C for 2\,hrs.
\label{RHall_anneal}}
\end{figure}

An interesting feature of GaMnAs semiconductors is that their
magnetic properties may be improved using low temperature
annealing \cite{Hayashi01, Potashnik01, KMYu02}. We have measured
the effect of annealing on the temperature dependence of the Hall
and bulk resistivities for three samples with $x$ equal 0.072,
0.085, and 0.087 which show an increase of the Curie temperature
of 21\,K, 25\,K, and 26\,K correspondingly upon annealing at
$260^{\circ}$C for 2\,hrs. The measured data and their fits for
$x=0.072$ are shown in Fig.~\ref{RHall_anneal}. These samples with
high Mn concentration have large resistivities (up to 632\,$\mu
\Omega \mathrm{m}$ at 300\,K for as grown $x=0.085$) and their Hall
coefficients at 300\,K are roughly 5 times bigger than for the
samples with $x\le 0.055$. For all measured samples, the bulk
resistivity and Hall coefficient decrease upon annealing (except
at the temperatures close to $T_c$ where Curie-Weiss susceptibilty
diverges). Fitting to the measured data for $x=0.072$ reveals an
approximately 100\% increase of the carrier concentration due to
annealing, which naturally contributes to the decrease of the bulk
resistivity.

\begin{figure}[h]
\includegraphics[width=3.5in]{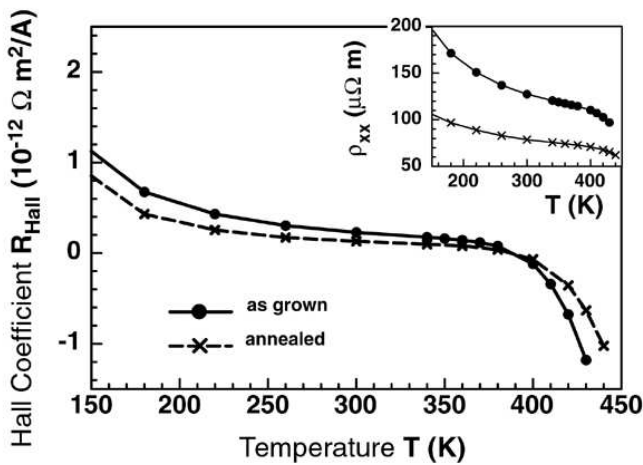}
\caption[RHall_drop]{High temperature behavior of the Hall
coefficient and bulk resistivity for $x=0.072$. Symbols are
measured data, lines are guides for an eye. \label{drop}}
\end{figure}

At  temperatures above 350--400\,K all studied samples show a
simultaneous drop in the Hall coefficient and bulk resistivity. At
sufficiently high temperatures the Hall coefficient changes sign
and increases rapidly in magnitude. The effect is reproducible for
different temperature scans. An example for $x=0.072$ is displayed
in the Fig.~\ref{drop}.
The samples with lower Mn concentration (3--5\%) exhibit the drop
at higher temperature, which looks qualitatively the same as in
Fig.~\ref{drop} but is shifted $\sim 50$\,K to the right. This
feature limited the fitting range of our calculations below 380\,K
(below 340\,K for the sample in Fig.~\ref{drop}). In order to check
the possibility of a thermally activated parallel conductance
through the material underneath the GaMnAs film, we measured the
bulk resistance of a bare GaAs substarate and Low-T MBE-deposited
GaAs film on a GaAs substrate used as a buffer layer for GaMnAs.
The resistances found at 400\,K are 4 orders of magnitude bigger
than the resistances of GaMnAs films which makes unlikely the
possibilty of the parallel conductance through the bulk GaAs.
However, the conductance may be due to an inversion layer formed
at the interface between GaMnAs film and GaAs buffer. A similar
effect of the resistivity drop and the sign change of the
magnetoresistance at high temperatures was observed and studied
for thin films on Si substrates and explained by the formation of
the inversion layer on the film/Si interface\cite{Tang02, Dai00}.
This phenomenon may limit the utility of our technique for samples
with higher $T_c$'s which would require extending the fitting
range to higher temperatures where the parallel conductance
through an inversion layer or substrate becomes appreciable.

\section{conclusions}

The results of the Hall effect measurements of a series of GaMnAs
samples above the Curie temperature and up to 380\,K can be described with a model equation which takes into account the
ordinary and anomalous contributions to the Hall resistivity. The good
agreement between the measured and fitting Hall data suggests that the
model correctly captures the physical origin of the Hall effect above
$T_c$ and up to 380\,K. We found that the anomalous
contribution to the Hall resistivity dominates over ordinary
contribution up to 380\,K in all samples.
  The fitting procedure based on
our model can be used as a new technique to determine the hole
concentrations in GaMnAs, which are notoriously difficult to measure
by conventional methods. The hole densities determined by our method
agree with independent measurements of carrier concentrations in our
GaMnAs samples done by means of ECV profiling.

The results of the analysis of the measured data based on our model
allow to discriminate between different existing theoretical approaches to the
anomalous Hall effect. Particularly, our observations of the AHE are
more consistent with a recently proposed theory of the Hall
conductivity in GaMnAs by Jungwirth {\em et al}\cite{Jungwirth02,
Jungwirth03} than they are with conventional theory which ascribes the
AHE to be due to the spin dependent scattering such as skew scattering
or side jump\cite{Chien}.

It is found that the temperature dependence of the anomalous Hall
effect above $T_c$ can be described with the Curie-Weiss law for the
paramagnetic susceptibility with the inclusion of a small, negative,
temperature and Mn content independent correction to the
susceptibility. The origin of this correction is not clear. It may be
due to the diamagnetic contribution from the GaAs matrix, but demonstrating
this will require an explanation for its anomalous size compared
to the paramagnetic contribution.

The Hall coefficient and bulk resistivity of all studied samples
exibit a drop above 400\,K.  The reason of such a dramatic change of
the transport parameters at high temperatures is not understood yet
but  may be caused by parallel conductance through an inversion layer on
the interface between the substrate and the GaMnAs film. This
penomenon complicates the study  of the transport properties of these
materials above 400\,K.

\begin{acknowledgments}
The authors gratefully acknowledge discussions with A. C. Ehrlich,
J. Kikkawa,  A. H. MacDonald, and T. Jungwirth. This work was
supported by the Office of Naval Research and the Research
Foundation for the State University of New York under grant number
N000140010951 and by the 21st Century Science and Technology Fund
of the State of Indiana.
\end{acknowledgments}


\begin{thebibliography}{99}

\bibitem{edmonds}  K. W. Edmonds, K. Y. Wang, R. P. Campion, A. C.
Neumann, N. R. S. Farley, B. L. Gallagher, C. T. Foxon, Appl. Phys. Lett. {\bf 81}, 4991 (2002).

\bibitem{y_ohno99}    Y. Ohno, D. K. Young, B. Beschoten, F.
Matsukura, H. Ohno, D. D. Awschalom, Nature (London) \textbf{402}, 790
(1999).

\bibitem{dietl01} T. Dietl, H. Ohno, F. Matsukura, Phys. Rev. B
\textbf{63}, 195205 (2001); T. Dietl, H. Ohno, F. Matsukura, J.
Cibert, D. Ferrand, Science \textbf{287}, 1019 (2000).

\bibitem{konig01} J. K\"{o}nig, J. Schliemann, T. Jungwirth, A. H.
MacDonald, cond-mat/0111314, in {\it Electronic Structure and Magnetism
of Complex Materials}, ed. D. J. Singh and D. A. Papaconstantopoulos, 
(Springer-Verlag, 2002).

\bibitem{h_ohno00} H. Ohno, D. Chiba, F. Matsukura, T. Omiya, E. Abe, T. Dietl, 
Y. Ohno, K. Ohtani, Nature (London) \textbf{408}, 944 (2000).

\bibitem{Koshihara} S. Koshihara, A. Oiwa, M. Hirasawa, S.
Katsumoto, Y. Iye, C. Urano, H. Takagi, H. Munekata, Phys. Rev. Lett. \textbf{78}, 4617
(1997).

\bibitem{Satoh01} Y. Satoh, D. Okazawa, A. Nagashima, J. Yoshino,
Physica E \textbf{10},  196 (2001).

\bibitem{Berciu01} M. Berciu, R. N. Bhatt, Phys. Rev. Lett.
\textbf{87} 107203 (2001).

\bibitem{DasSarma03} S. Das Sarma, E. H. Hwang, A. Kaminski,
Phys. Rev. B \textbf{67} 155201 (2003).

\bibitem{omiya} T. Omiya, F. Matsukura, T. Dietl, Y. Ohno, T. Sakon, M. Motokawa, H. Ohno, 
Physica E \textbf{7}, 976 (2000).

\bibitem{baxter02} D. V. Baxter, D. Ruzmetov, J. Scherschligt, Y.
Sasaki, X. Liu, J. K. Furdyna, C. H. Mielke, Phys. Rev. B \textbf{65}, 212407 (2002).

\bibitem{Yu02} K. M. Yu, W. Walukiewicz, T. Wojtowicz, W. L. Lim, X. Liu, Y. Sasaki, 
M. Dobrowolska, J. K. Furdyna, Appl. Phys. Lett. \textbf{81}, 844 (2002).

\bibitem{Seong02} M. J. Seong, S. H. Chun, H. M. Cheong, N.
Samarth, A. Mascarenhas, Phys. Rev. B \textbf{66}, 033202 (2002).

\bibitem{Jungwirth02} T. Jungwirth, Q. Niu, A. H. MacDonald, Phys.
Rev. Lett. \textbf{88}, 207208 (2002).

\bibitem{Jungwirth03} T. Jungwirth, J. Sinova, K. Y. Wang, K. W. Edmonds, 
R.~P.~Campion, B. L. Gallagher, C. T. Foxon, Q. Niu, A. H. MacDonald, 
Appl. Phys. Lett. {\bf 83}, 320 (2003).

\bibitem{Chien} \textit{The Hall Effect and Its Applications}, edited
by C. L. Chien and C. R. Westgate (Plenum, New York, 1980), pp.
43-51, 56-67.

\bibitem{Martin1} D. H. Martin, \textit{Magnetism in Solids}, (The
MIT press, Cambridge, Massachusetts, 1967), p. 20.

\bibitem{H.Ohno99} H. Ohno, J. Magn. Magn. Mater. \textbf{200}, 110
(1999).

\bibitem{Faktor} M. M. Faktor, T. Ambridge, C. R. Elliott, J. C. Regnault in {\it Current Topics in
Materials Science},  (North Holland, Amsterdam, 1980), Vol. 6, ed. E. Kaldis, pp. 1-107;
P.~Blood, Semicond. Sci. Technol. \textbf{1}, 7 (1986).

\bibitem{Hayashi01} T. Hayashi, Y. Hashimoto, S. Katsumoto, Y.
Iye, Appl. Phys. Lett. \textbf{78}, 1691 (2001).

\bibitem{Potashnik01} S. J. Potashnik, K. C. Ku, S. H. Chun, J. J.
Berry, N.~Samarth, P. Schiffer, Appl. Phys. Lett \textbf{79}, 1495 (2001).

\bibitem{KMYu02} K. M. Yu, W. Walukiewicz, T. Wojtowicz, I.
Kuryliszyn, X. Liu, Y. Sasaki, J. K. Furdyna,   Phys. Rev. B \textbf{65}, 201303(R)
(2002).

\bibitem{Tang02} J. K. Tang, J. B. Dai, K. Y. Wang, W. L. Zhou, N. Ruzycki, U. Diebold,
 J. Appl. Phys. \textbf{91}, 8411 (2002).

\bibitem{Dai00} J. Dai, L. Spinu, K. Y. Wang, L. Malkinski, J.
Tang, J. Phys. D \textbf{33}, L65 (2000).

\end{thebibliography}

\end{document}